\documentstyle[aas2pp4,tighten]{article}
\def\nice#1::::{#1}    \def\subm#1::::{}   % format ``pretty print''
\nice \def\apjsmall{\small} ::::
\subm \def\apjsmall{} ::::
\nice \input epsf.sty ::::

\nice \textheight 260mm \headheight 0pt \topmargin -2.1cm ::::

\def\k{\,km\,s$^{-1}$}
\def\centreline{\centerline}
\def\.{{\cdot}} 
\def\gtapprox{\,\lower.6ex\hbox{$\buildrel >\over \sim$} \, }
\def\ltapprox{\,\lower.6ex\hbox{$\buildrel <\over \sim$} \, }
\def\propapprox{\,\lower.6ex\hbox{$\buildrel \propto\over \sim$} \, }

\def\arcs{\ifmmode {'' }\else $'' $\fi}     %Arc seconds%
\def\arcm{\ifmmode {' }\else $' $\fi}     %Arc minutes%
\def\deg{\ifmmode^\circ\else$^\circ$\fi}    %Degree sign%

\def\fr7{7$ \hskip -0.9ex \vrule height0.8ex width0.8ex depth-0.73ex
                                                                \hskip0.1ex$}
\def\frtoday{le\space\number\day\space\ifcase\month\or
  janvier\or f\'evrier\or mars\or avril\or mai\or juin\or
  juillet\or ao\^ut\or septembre\or octobre\or novembre\or d\'ecembre\fi\space \number\year}

\newcommand\joref[5]{#1, #5, {#2, }{#3, } #4}  % journal reference
\newcommand\confref[5]{#1, #5, {#2, }{#3, } #4.} % conference reference
\newcommand\inpress[5]{#1, #5, #2, in press}  % journal reference
\newcommand\epref[3]{#1, #3. #2}
\def\DCP{DCP} % Change this if we choose a different term!

\begin{document}

%\nice  \epsfbox[155 578 462 579]{"draft.ps"} ::::

\title{A Possible Effect of the Period of Galaxy Formation on 
the Angular Correlation Function}
\author{Tomoya Ogawa}
\affil{Graduate School of Science and Technology, 
Chiba University, Inage-ku, Chiba 263, Japan}
\and
\author{Boudewijn F. Roukema}
\affil{Division of Theoretical Astrophysics, National Astronomical 
Observatory, Mitaka, Tokyo 181, Japan}
\and
\author{Kazuyuki Yamashita}
\affil{Information Processing Center, Chiba University, Inage-ku,
Chiba 263, Japan}

\author{Version: \frtoday}

\begin{abstract}
During the epoch of galaxy formation, 
the formation of the first galaxies in regions of high overdensity 
may lead to an initial 
amplitude of the spatial correlation function of galaxies, $\xi_0,$ 
much higher than that expected in linear perturbation theory. This 
initially high ``bias'' would consequently decrease to the near-unity
values expected from local observations. 
Such a ``Decreasing Correlation Period'' ({\DCP}), 
detected in N-body simulations
under certain conditions by several authors, is parametrised 
in a simple way, ending at a ``transition'' redshift $z_t.$ 

Consequences 
on $w_0,$ the amplitude of the angular correlation
function of faint galaxies (for a fixed apparent magnitude range), 
have been estimated. The sensitivity of $w_0$
to $z_t,$ and how this is affected by 
the redshift distribution of the galaxies in the sample,
by low redshift spatial correlation function behaviour, by cosmology, and
by the rate at which $\xi_0$ decreases during the {\DCP}, are 
shown and discussed.
This simple model is intended as a first exploration of
the effects of the {\DCP} for a likely range of 
parameter space. 

Comparison with the Hubble Deep Field (HDF) 
estimate of $w_0$ (\cite{Vill96}~1996)
indicates that the {\DCP} is compatible with observation for 
values of $z_t$ in the theoretically expected range 
$z_t \gtapprox 1.$
However, tighter observational 
constraints on the HDF redshift distribution and 
on low redshift correlation function growth will be needed either to 
detect the {\DCP} unambiguously or to constrain $z_t$ to redshifts greater
than those of galaxies in existing surveys.

\end{abstract}


%%%%%%%%%%%%%%%%%%%%%%%%%%%%%%%%%%%%%%%%%%%%%%%%%%%%%%%%%%%%%%%%%%%
%% Figure section. Defined early so that position in output can
%% be easily moved towards earlier pages if LaTeX wants to put
%% them all at the end...
%%%%%%%%%%%%%%%%%%%%%%%%%%%%%%%%%%%%%%%%%%%%%%%%%%%%%%%%%%%%%%%%%%%
\def\fschem{ \begin{figure}[h]
\centering 
\nice \centreline{\epsfxsize=7cm
 \epsfbox[18 144 592 718]{"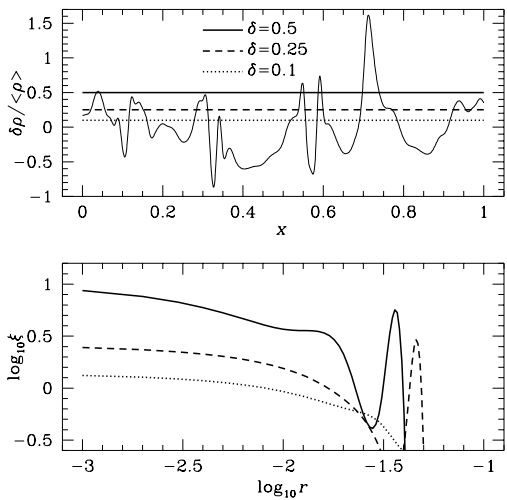"}} ::::
\caption[Neg corrn growth-schematic]{ \apjsmall
\label{f-schem} 
Semi-schematic demonstration of how $\xi_0$ may 
{\em decrease} during the galaxy formation epoch (as time increases). 
The upper panel shows a fluctuation profile of gas density 
$\rho/\left\langle \rho \right\rangle$ in a one-dimensional 
``core sample'' through KY's simulation (at $z=9$), 
with ``galaxies'' 
approaching non-linear collapse occurring in the regions above 
$\delta\equiv \delta\rho/\left\langle \rho \right\rangle=0\.5$ 
(solid line). Dashed and dotted lines are for lower density (``linear'') regions.
The lower panel shows one-dimensional $\log_{10}\xi$ for ``collapsed'' galaxies 
(solid line) and ``future'' galaxies (dashed, dotted lines) against separation. 
The length unit is 20$h^{-1}$~Mpc (comoving). 
This figure is intended as illustration, 
not for precise quantitative estimates.}
\end{figure} }

\def\fcosmo{ \begin{figure}
\nice 
\epsfxsize=7cm
 \epsfbox[18 144 592 718]{"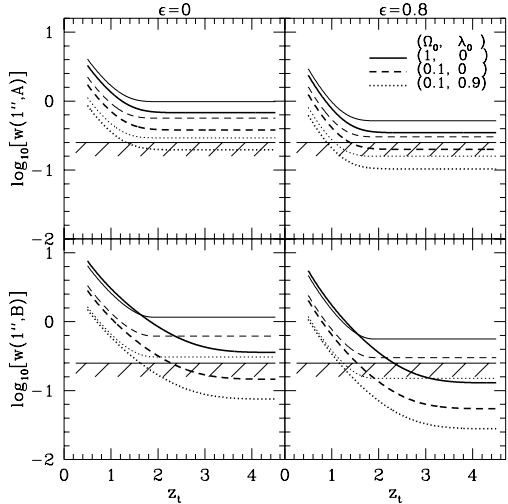"} 
 ::::
\caption[Cosmo Figure]{ \apjsmall
\label{f-cosmo} Effects of $z_t$ (Eqn~\protect\ref{e-eps_nu}) 
on $\log_{10}(w_0),$ where $w_0\equiv w(1\arcs,\partial N/\partial z).$ 
Redshift distributions are 
peaked around $z_0=1\.5$ (A) in the upper panels and $z_0=3$ (B) in 
the lower panels. 
Lefthand panels are for clustering stable in proper coordinates; righthand 
panels are for linear perturbation growth. 
Cosmological geometries 
are shown as solid ($\Omega_0=1, \lambda_0=0$), 
dashed ($\Omega_0=0\.1, \lambda_0=0$) and 
dotted ($\Omega_0=0\.1, \lambda_0=0\.9$) curves. Curves are 
thick for $z_c=\infty$ and 
thin for $z_c=2.$ All panels are for $\nu=2.$ 
Shading below $\log_{10}(w_0)=-0\.6$ 
shows a $1\sigma$ upper limit to the HDF estimate of $w_0$ for $27<R<29$ 
(\protect\cite{Vill96}~1996). 
}
\end{figure} 
} %end of \def\fcosmo

\def\fnu{ \begin{figure}
\nice 
\epsfxsize=7cm
 \epsfbox[18 144 592 718]{"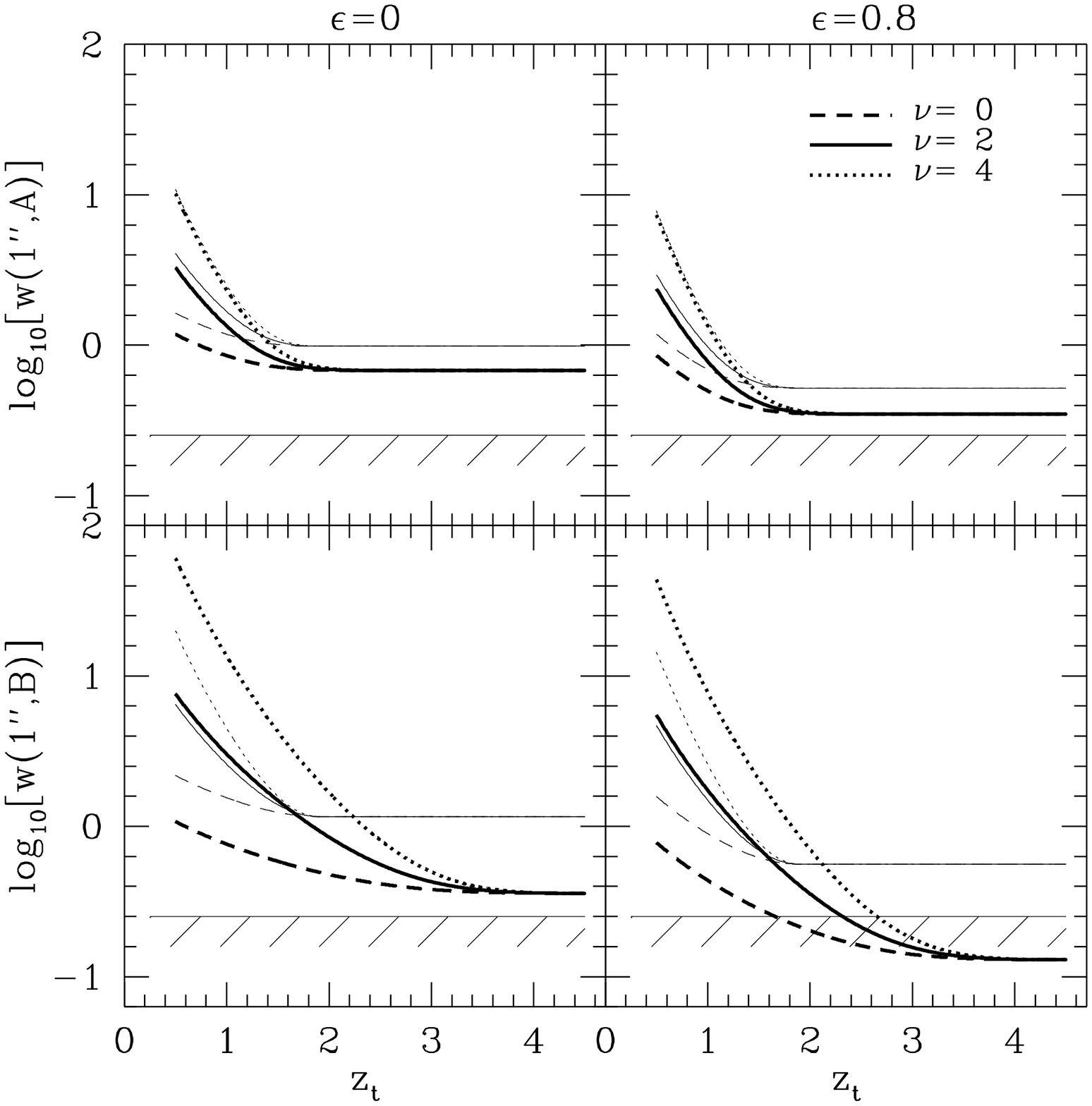"} 
%\vspace{7cm}
 ::::
\caption[$\nu$ Figure]{ \apjsmall
\label{f-nu} Sensitivity of $w_0$ to $\nu$ for an EdS geometry. 
Differing $\nu$ is shown by dashed ($\nu=0$), solid ($\nu=2$)
and dotted ($\nu=4$) curves.
The panels are otherwise identical to
those of Fig.~\protect\ref{f-cosmo}, except for a small 
offset in the vertical range.
}
\end{figure} 
} %end of \def\fcosmo

%%%%%%%%%%%%%%%%%%%%%%%%%%%%%%%%%%%%%%%%%%%%%%%%%%%%%%%%%%%%%%%%%%%

\section{INTRODUCTION}
Present-day galaxies---or their underlying dark matter haloes---are usually 
considered to be discrete objects, since their densities are 
``non-linear'' (much greater than the average matter density). 
Effective discreteness, 
due to non-linearity, may play a role on larger length scales. 
During the period of galaxy formation, 
many small length-scale perturbations superimposed on 
high amplitude large length-scale perturbations should have their 
overdensities boosted into the non-linear regime, whilst 
small scale perturbations in other regions 
remain linear. Such (temporary) restriction of halo (or galaxy) formation to 
discrete regions may overwhelm the spatial correlation function, $\xi,$ expected 
from linear fluctuation theory, so that the evolution of 
$\xi$ for haloes (or galaxies) during this period may be very different 
to the linear expectation. 
High power on large length scales can thus lead to an effective discreteness 
which causes a temporary period of ``biassed'' galaxy formation. 
Other regionally-dependent galaxy formation criteria (e.g., a gas 
cooling condition) could similarly cause ``biassed'' galaxy formation. 
\nice\fschem::::
\placefigure{f-schem}

If the period of such conditional galaxy formation 
continues to observable epochs, 
then existing observations may reveal such an effect. 
The purpose of this article is to alert the reader to the 
degree to which this effect could be important in the interpretation 
of observations of the normalisation of the faint galaxy angular 
correlation function, $w_0\equiv w(\theta,m),$ where $\theta$ and $m$ are 
a fixed angle and apparent magnitude respectively. 
A more detailed motivation is given in \S\ref{s-dcp}, 
a simple model for comparison to observation is presented in \S\ref{s-models} 
and results are shown in \S\ref{s-results}. 
The relevance of the results is discussed in \S\ref{s-discuss}
and \S\ref{s-conclu} concludes. 

\section{THE DECREASING CORRELATION PERIOD}\label{s-dcp}
If the slope, $n,$ of the initial perturbation spectrum on galaxy scales 
is similar to that in the CDM model ($n\approx-2$), then 
there is relatively more ``power'' on large (length) scales than 
on small scales. 
Hence, the first 
collapsed objects---haloes---should 
appear within isolated ``clusters'' in the 
high density large length-scale perturbations (cf. Fig.~\ref{f-schem}). 
The lack of 
haloes between these ``clusters'' lowers 
the mean halo number density.
Since the normalisation, $\xi_0\equiv \xi(r_0,z)$  
(see Eqn~\ref{e-eps_nu}), of 
$\xi$ for haloes (or galaxies) depends on 
the mean halo (or galaxy) number density, 
the objects in the ``clusters'' yield high values of $\xi_0,$ 
i.e., a very high initial bias, 
$b\equiv \xi_{\mbox{\rm \small halo}}/\xi_{\mbox{\rm \small matter}}.$ 
As objects in other 
regions successively reach the turnaround density 
($\delta(\rho)/\rho \approx 5\.6$, \cite{Kih68}~1968; 
\cite{Bible80}~1980, Eqn~19.50) 
$b$ successively 
decreases to the near-unity values expected from present-day galaxy 
observations. (For a recent review of low $z$ estimates of $\xi$ and 
$b,$ see \cite{Peac96}~1996.) 

Such a period during which $\xi_0$ may decrease 
is hereafter referred to as the {\DCP} (Decreasing Correlation Period). 
A {\DCP}---for dark matter haloes---has been 
found in pure gravity N-body simulations 
for a power law initial perturbation spectrum of $n=-2$ 
(\cite{Rouk93}~1993; \cite{RQPR97}~1997) 
and for a CDM perturbation spectrum 
(\cite{BV94}~1994). 
Both cases are for 
an $\Omega_0=1, \lambda_0=0$ cosmology. The authors find that 
the {\DCP} finishes at $z\gtapprox 1.$ 
A {\DCP} has also been found---for galaxies---in 
cosmological N-body simulations including both gravity and hydrodynamics, 
for a CDM perturbation spectrum 
and $\Omega_0=0\.2, \lambda_0=0$ 
(unpublished simulations by one of us, KY, 
similar to those of \cite{Yama93}~1993). 
The {\DCP} finishes 
at similar redshifts in this case. 

The {\DCP} requires more power on large than small length scales, 
(unless another regionally-dependent galaxy formation condition is adopted). 
This is clearly seen in \cite{RQPR97}'s analysis: the {\DCP} was not 
found in the simulations with $n=0.$ 

The existence of the {\DCP} also seems sensitive to halo detection conditions. 
\cite{BV94}~(1994) found the {\DCP} for an 
overdensity detection threshold of $\sim250$ but not for a threshold of $\sim2000.$ 

Rather than making specific galaxy 
formation assumptions in order to get a precise 
quantitative prediction 
of the behaviour of $\xi_0$ during the {\DCP}, 
the approach chosen is a simple parametrisation of $\xi_0$ during 
this period and an exploration of 
a likely region of parameter space. 
Sensitivity of $w_0$ to the ``normal'' growth rate of structure with 
decreasing redshift, to the redshift distribution of galaxy samples, 
and to cosmology are examined. 

\section{MODELS} \label{s-models}
We model the {\DCP} behaviour of $\xi$ as follows. The {\DCP} starts at 
$z_c$ (high $z$ ``cutoff'') and ends at $z_t$ (``transition'' to 
normal $\xi_0$ behaviour), 
such that 
\begin{equation}
\xi(r,z)=\left\{ 
	\begin{array}{ll}
	\left[(1+z)\over(1+z_t)\right]^\nu \xi(r,z_t),
		& z_c > z > z_t \\
	(r_0/r)^\gamma (1+z)^{-(3+\epsilon-\gamma)}, & z_t \ge z >0
	\end{array}
	\right.
\label{e-eps_nu}
\end{equation}
where $r$ is in comoving coordinates, 
$r_0=5\.5 h^{-1}$~Mpc and 
$\gamma = 1\.8$ are (respectively) the 
correlation length and 
(negative of) the slope of the 
local galaxy correlation function, 
$\epsilon$ parametrises the low 
redshift (``unbiassed'') growth of $\xi_0$ 
(e.g., \cite{GroP77}~1977) and 
 $\nu$ parametrises the 
rate at which $\xi_0$ 
decreases during the {\DCP}.

As indicated above, $z_t\gtapprox 1$ is expected, 
while $\nu$ could be as 
low as $0\.3$ (\cite{Rouk93}~1993; \cite{RQPR97}~1997) or as high as $4\.5$ 
(KY's simulation); hence the ranges explored here. 

The angular correlation function was chosen for this 
article because it 
constitutes an observationally feasible (but indirect) way of estimating 
$\xi_0$ for $z>1.$ 
The link between the two functions, 
described by Limber's equation (e.g., 
\cite{Ph78}~1978), involves the redshift distribution, 
${\partial N \over \partial z}(z,m)$ for a nominal $m$ that
implicitly depends on surface brightness limit, 
minimum detection diameter and seeing conditions. 

We examine the extent to which the {\DCP} could be detected
in $w_0$ observations. 
While $w_0$ for 
brighter magnitudes (e.g., \cite{Ef91}~1991, 
\cite{Neu}~1991, \cite{RP94}~1994, \cite{BSM95}~1995) 
might include galaxies at high enough redshifts for the {\DCP} 
to be important, 
the Hubble Deep Field (HDF, \cite{Will96}~1996) 
estimate of $w_0\equiv$ $w(1\arcs,$ $27\ltapprox R\ltapprox 29)$ 
for faint galaxies 
(\cite{Vill96}~1996) is considered here, 
since the HDF almost certainly contains many galaxies at 
$z>1$ (e.g., \cite{GwHart96}~1996). Assumptions on cosmology, 
galaxy formation and galaxy evolution are required for precise derivation 
of ``photometric'' redshifts, so instead, 
simple analytical 
redshift distributions of the form 
\begin{equation}
{\partial N /\partial z}
\propto z^2 \exp[-(z/z_0)^\beta] \label{e-dndz} 
\end{equation} are adopted 
(after \cite{Ef91}~1991; \cite{Vill96}~1996), 
with \cite{Vill96}'s value for the 
high-$z$ cutoff parameter, $\beta=2\.5,$ 
and ``characteristic'' redshifts which bracket a realistic range: 
(A) $z_0=1\.5$  and (B) $z_0=3.$ 

Theoretically motivated values of $\epsilon$ include 
$\epsilon=-1\.2$ for clustering fixed in comoving 
coordinates (for $\gamma=1\.8$); 
$\epsilon=0$ for clustering fixed in proper coordinates; and 
$\epsilon=0\.8$ for $\xi_0$ of linearly 
growing matter density perturbations in an Einstein-de Sitter (EdS) cosmology. 
The best direct estimates of $\epsilon$ for $z<1$ 
indicate $0< \epsilon < +2$ (\cite{CFRS-VIII}~1996), so 
$\epsilon=0$ and $\epsilon=0\.8$ are adopted.

Since $w_0$ probes both galaxy formation and cosmology, 
EdS, open and non-zero cosmological constant 
geometries, with the parameters indicated in 
Figure~\ref{f-cosmo}, are investigated. 

\nice\fcosmo::::
\placefigure{f-cosmo}

\nice\fnu::::
\placefigure{f-nu}

\section{RESULTS} \label{s-results}

Figures~\ref{f-cosmo} and \ref{f-nu} show the effects of a {\DCP} on 
$w_0$. The following summary of the results 
mostly applies for $z_c=\infty$ 
(no $z$ cutoff, thick curves). 
Note that for small $z_c,$ 
the number of high $z$ galaxies cut off by $z_c$ is much greater for 
a redshift distribution with high-$z_0$ than for one with low-$z_0.$ 
Hence, $z_c$'s effect is stronger for the former than 
for the latter: the difference between the 
$z_c=\infty$ and $z_c=2$ (thin) curves is much bigger in the 
lower panels than in the upper ones. 

(1) The curves are most strikingly affected by 
$z_t.$ For $z_t \gg z_0,$ i.e., if the {\DCP} finishes by a redshift 
greater than typical redshifts 
in ${\partial N / \partial z},$ then 
it cannot have an effect on $w_0,$ so the curves are flat. As $z_t$ drops 
below $z_0$, 
the ``initially high bias'' starts to have an effect on $w_0,$ and becomes 
stronger as $z_0-z_t$ increases. This simply means 
that for a given redshift distribution and a given $\epsilon,$ 
as we suppose that the {\DCP} finishes at 
later epochs, the (fixed) galaxy redshifts 
become more representative of the {\DCP}. Since $\nu$ is held 
constant, $\xi_0$ values increase. 
Hence, $w_0$ increases. 

(2) 
In the absence of the DCP, or equivalently, if $z_t \gg z_0$, a higher 
$z_0$ would imply a lower $w_0$ (for a fixed $\epsilon$),
since $\xi_0$ grows with time. 
However, for a fixed $z_t,$ the {\DCP} 
causes $w_0$ to be sensitive to $z_0$ 
in a way opposite to that 
expected in the absence of the {\DCP}. 
As $z_t$ decreases to well below $z_0,$ the effect 
of the {\DCP} becomes strong enough to counter the effect of normal 
$\xi_0$ growth, so that 
a {\em lower} $z_0$ implies a lower $w_0$. 
In other words, $\xi_0$ has a minimum at $z=z_t,$ 
so $w_0$ is lowest for $z_0 \approx z_t.$

The effect of $z_0$ on $w_0,$ in the presence of 
a {\DCP}, could be useful for constraining $z_t.$ The 
higher the value of $z_0,$ the more 
sensitive $w_0$ would be to $z_t.$ Observations of $w_0$ for the higher 
$z$ galaxies should be better than estimates based on lower $z$ 
galaxies.

(3) The effects of cosmological geometry on $w_0$ are similar to 
those found by others (e.g., \cite{YTP93}~1993). 
If the $z_0$ and $\epsilon$ values adopted here are accepted as realistic 
for the faint HDF galaxies, and if the $w_0$ estimate is accepted as 
unbiassed by systematic effects (cf. \cite{Colley96}~1996), 
then an EdS 
universe could only be consistent (at $1\sigma$) with the observation 
if $z_0 \gtapprox 3,$ $\epsilon \gtapprox 0\.8$ and $z_t \gtapprox 2.$
Larger values of $z_0$ or $\epsilon$ would allow lower $z_t$.

On the other hand, for a $\Lambda$-dominated flat geometry, 
the HDF $w_0$ would be consistent with any of the 
$z_0$ and $\epsilon$ values adopted, provided that $z_t \gtapprox 1\.5.$ Hence,
the constraint on $z_t$ is much weaker for such a geometry
than for the EdS geometry. An open geometry gives an intermediate
constraint.

(4) Increasing $\epsilon$ results in lower values of $\xi_0$ at 
low redshifts (since $\xi_0$ is normalised at $z=0$), and 
via continuity at $z=z_t$ also gives lower $\xi_0$ 
during the {\DCP}. Hence, increasing $\epsilon$ generally results in
lower $w_0.$ 
Moreover, in the region of parameter space where the {\DCP} has an effect, 
i.e., for $z_t \ltapprox z_0,$ a higher $\epsilon$ for a fixed 
$z_0$ gives higher sensitivity to $z_t.$ However, 
since we use a constant $\nu,$ the converse does not hold: 
 for $z_t \ltapprox z_0,$ increasing $\epsilon$ for a fixed 
$z_t$ does not change the sensitivity of $w_0$ to $z_0.$ 
These dependencies can be seen by careful examination of Figures~\ref{f-cosmo} 
and \ref{f-nu}. 

(5) As seen in Figure~\ref{f-nu},
for fixed $z_0$ and $\epsilon,$ $\nu$ has little effect 
if $z_t \gtapprox z_0,$ since very few galaxies are located in the {\DCP}; 
but if $z_t < z_0,$ higher $\nu$ implies higher 
$\xi_0$ values, hence, higher $w_0.$ 

Given the region of parameter space presented in 
Figure~\ref{f-nu} (and the HDF observation), 
most predicted values of $w_0$ are too high, so 
the only advantage of a smaller $\nu$ would be to slightly relax the 
EdS constraint on $z_t$ from $z_t \gtapprox 2$ to $z_t \gtapprox 1\.5.$ 

However, if $z_0$ or $\epsilon$ are higher than considered 
here, 
or if the Universe is $\Lambda$-dominated, 
then the value 
of $\nu$ would be more important. It should then be noted that 
$w_0$ is more sensitive 
to $z_t$ for higher values of $\nu$; and for a fixed $z_t$ and $\epsilon,$ 
increasing $z_0$ increases the sensitivity of $w_0$ to $\nu.$ 

\section{DISCUSSION} \label{s-discuss}
What is the relevance of the above effects given present 
estimates of $z_0$ and $\epsilon$?
Our present ignorance of the redshift distribution of
the galaxies in the HDF (at $27<R<29$) and uncertainty regarding 
the spatial correlation function 
behaviour at redshifts lower than $z_t$ mean that the constraints
on $z_t$ presented above should not be considered definitive.

Different authors using 
broad-band ``photometric'' redshift estimates of the HDF galaxies 
find that $z_{\mbox{\rm \small median}}=2\.1$
(\cite{Mob96}~1996), 
$z\approx 2$ (\cite{Metc96}~1996), 
or that the redshift distribution is bimodal, with peaks at
$z=0\.6$ and $z=2\.2$ (\cite{GwHart96}~1996). 
These median values are bracketted
by our canonical values of $z_0=1\.5$ (A) and $z_0=3$ (B) (except for
\cite{GwHart96}'s low redshift component), so our estimate of $z_t$ 
would probably only change a little using more accurate redshift 
distributions.

More of concern is the low redshift spatial correlation
growth, parametrised by $\epsilon$. Although the CFRS estimate 
(\cite{CFRS-VIII}~1996), based on galaxies with 
$z_{\mbox{\rm \small median}}=0\.56,$ 
can be summarised by stating that $0<\epsilon<2,$
use of the correlation length of $r_0=5\.0 h^{-1}$~Mpc and the 
median redshift ($cz_{\mbox{\rm \small median}}=15,200$\k) 
of the ``zero redshift'' Stromlo-APM survey (\cite{Love92}~1992, 1995) 
would imply that $\epsilon=2\.8$ (cf. \S 4.1(3), \cite{CFRS-VIII}). 
Measurements at lower median redshifts indicate that 
$\epsilon=1\.6\pm0\.5$ (\cite{WIHS93}~1993, 
$z_{\mbox{\rm \small median}}=0\.4$), 
$\epsilon=-2\.0\pm2\.7$ 
(\cite{CEBC}~1994, $z_{\mbox{\rm \small median}}=0\.16$)\footnote{Note that 
$\varepsilon$ used by \protect\cite{CEBC} relates to
$\epsilon$ by $\epsilon=\gamma( 1- \varepsilon ) -3.$}
or that $\epsilon=0\.8^{+1.0}_{-1.3}$ 
(\cite{Shep96}~1996, $z_{\mbox{\rm \small median}}=0\.36$). The higher of
these values would imply that the {\DCP} could be present to redshifts
lower than unity without violating the HDF $w_0$ estimate.

\section{CONCLUSIONS}\label {s-conclu}
Consequences of a possible {\DCP} on $w_0$ 
have been presented for a simple parametrisation of $\xi.$ 

For a fixed redshift distribution and $\epsilon,$ 
the expected value of  $w_0$ can be an order of magnitude or 
more higher than that expected in the absence of a {\DCP}. The 
sensitivities of $w_0$ to $z_t,$ $z_0,$ $\epsilon,$ 
$\Omega_0,$ $\lambda_0$ and $\nu$ have been presented. 

For an EdS geometry, the \cite{Vill96}~(1996) HDF estimate 
of $w_0$ constrains $z_t$ to $z_t \gtapprox 1\.5$ 
over the adopted parameter range, 
requiring that $z_0\gtapprox 3$ and 
$\epsilon\gtapprox 0\.8.$ This estimate of $z_t$ is 
consistent with the theoretically 
expected $z_t \gtapprox 1.$ 
For $\Omega_0=0\.1, \lambda_0=0\.9,$ $z_t \gtapprox 1$ would be 
compatible with a much wider region of parameter space. 

Of course, these estimates assume that the region of parameter
space covered here is sufficiently large to include the real values.
Lack of spectroscopic redshifts and observational indications that 
$\epsilon$ might be much higher than considered here mean that the
{\DCP} might in fact be present to much lower redshifts.

In any case, it seems important that a possible {\DCP} should be considered 
when interpreting faint galaxy $w_0$ estimates. Moreover, since the {\DCP} 
is dependent on the power index $n,$ 
precise $w_0$ constraints on $z_t$ and $\nu$ 
might provide a new constraint on $n$ on relevant scales. 

\medskip
We thank S. Miyaji and R. Matsumoto for valuable suggestions. 
Use was made of computations performed on a VPP-500 (RIKEN) and 
on a CS6432 and a HITAC S-3800 (I.P.C., Chiba University). 
BFR acknowledges a Centre of Excellence 
Visiting Fellowship (N.A.O.J., Mitaka). 

\subm \clearpage ::::

%end

\subm \clearpage ::::

\subm \fschem ::::
\subm \fcosmo ::::
\subm \fnu ::::
%\subm \fexplan ::::

\end{document}